\documentclass[twocolumn,secnumarabic,amssymb,nobibnotes,aps,prd,showpacs]{{revtex4-1}}
\usepackage{amsmath}
\usepackage{amssymb}
\usepackage[dvipdfm]{epsfig}
\usepackage{graphicx}
\begin{document}
\title {Characteristics of Power Loss in SMC a Key for Desining the Best Values of
Technological Parameters 
}  
\author{Barbara \'{S}lusarek}
\email[e-mail: ]{barbara.slusarek@itr.org.pl}
\affiliation{Tele and Radio Research Institute, ul. Ratuszowa 11, 03-450 Warszawa, Poland}
\author{Bartosz Jankowski}
\email[e-mail: ]{bartosz.jankowski@itr.org.pl}
\affiliation{Tele and Radio Research Institute, ul. Ratuszowa 11, 03-450 Warszawa, Poland}
%\affiliation{Tele and Radio Research Institute, ul. Ratuszowa 11, 03-450 Warszawa, Poland}
\author{Krzysztof Sokalski}%
% ZMIANA podkreslenie
\email[e-mail: ]{sokalski\_krzysztof@o2.pl}
%\email[e-mail: ]{sokalski\_krzysztof@o2.pl}
% ZMIANA revtex sam podzieli linie
\affiliation{Institute of Computer Science, 
Cz\c{e}stochowa University of Technology,
Al. Armii Krajowej 17, 42-200 Cz\c{e}stochowa, Poland}
\author{Jan Szczyg{\l}owski}
\email[e-mail: ]{jszczyg@el.pcz.czest.pl}
\affiliation{Institute of Power Engineering, Cz\c{e}stochowa University of Technology, Al. Armii Krajowej 17, 42-200 Cz\c{e}stochowa, Poland} \affiliation{Eletrotechnical Institute, Ul. Pozaryskiego 28, 04-703 Warszawa, Poland} 

%\keywords{Soft magnetic composites; Core loss scaling; Energy loss characteristics; Technological parameters}

%\date{}
\begin{abstract}
Optimization of power loss in soft magnetic components basis on the choice of the best  technological parameters values. Therefore,
the power losses have been measured in Somaloy 500 samples for a wide range of frequency and magnetic induction. These samples have been prepared under a wide range of the hardening temperatures and pressures. The power loss characteristics have been derived by assuming that investigated samples obeyed the scaling law. Agreement obtained between experimental data and the scaling theory has confirmed this assumption. Moreover, the experimental data of the given sample have been collapsed to a single curve which represented  measurements for all values of frequency an magnetic induction pick. Therefore, the scaling transforms the losses characteristics from the two dimensional surfaces to the one dimensional curves. The samples were produced according two methods: for different pressures  with constant temperature  and at different temperatures with constant pressure. In both cases the power losses decrease with increasing pressure and with increasing temperature. These trends in decreasing losses stopped for certain critical values of pressure and temperature, respectively. Above these values the power losses increase suddenly. Therefore,  the mentioned above the critical pressure and the critical temperature are sought after solutions for optimal values. In order to reduce the parameters values set the limit curve in the  pressure-temperature plane has been derive. This curve constitutes a separation curve between the parameters values corresponding to high and low losses.
\end{abstract}
\pacs{44.05.+e,75.90.+w}
\maketitle
 \section{Introduction}\label{I}
 The expanding interest in the introduction of Fe-based composite materials, also called the Soft Magnetic Composite (SMC), in electrical devices, such as electrical motors, is in obvious ways connected with their properties. The principle feature of these materials is that iron particles are insulated by a thin organic or inorganic coating. SMC offers several advantages over laminated steel sheet, e.g. isotropic magnetic properties, the opportunity to tailor their physical properties, very low eddy current loss, relatively high resistivity, high magnetic permeability \citep{bib:Shokrollahi}.
Fe-based composite materials are used in the manufacture of electrical devices owing to their physical properties. The most common use of these materials is in various electrical motors, e.g. a high speed permanent magnet motor, a synchronous electric motor, an axial-flux permanent magnet synchronous machine, a claw pole permanent magnet motor \citep{bib:Wojciechowski}, \citep{bib:Hamlera}, \citep{bib:Stefano}, \citep{bib:Guo}. Furthermore, soft magnetic composites are used in an electromagnetic actuator \citep{bib:Jack} and in electromagnetic shielding \citep{bib:Kim}.
The manufacturing process of magnetic composites can be divided into a few stages. Every stage is characterized by a set of technological parameters, which may ultimately affect the outcome parameters of magnetic composite materials. Three of the most important processing considerations of compression moulding techniques are pressure, temperature and time of curing. The value of compaction pressure has a strong influence on the initial, 'green', and final compacted densities, which is reflected in some of the physical properties of magnetic composites. On the other hand, changes in temperature or time of curing lead to variations in the dielectric parameters of insulation layer. It is all the more relevant because inappropriate selection of these parameters can lead to an increase in total energy losses.
Development of electrical devices, such as electric motors or actuators, is closely connected with development of magnetic materials, obviously. However, the mere application of modern magnetic materials is no guarantee of success. This implies the need to look for new ways to design the best value of technological parameters. 
In the last decade some methods for determining these values have been elaborated by a few laboratories \citep{bib:Dobrz},\citep{bib:Dobrz2}, \citep{bib:Pang},\citep{bib:Gilbert},\citep{bib:Lefe},\citep{bib:Lemi},\citep{bib:Gelin}. Although none of them have formulated their method in the algorithmic form, that mainly one can distinguish the following common steps in the applied procedures:\\

\begin{itemize}
\item  The series of samples are prepared  of different powders compositions and processing methods. 
\item  The core loss versus magnetic induction and  frequency is measured. Usually  the results of these  measurements are presented as energy loss characteristics versus $B_{m}$ for the constant values of $f$.  
\item The particles content and the values of technological parameters corresponding to the lowest energy loss are reported for desired values of $B_{m}$ and $f$. 
\end{itemize}
The aim of this paper is to work out systematic procedures for designing the best values of
technological parameters and optimizing the power losses.  Realization of this venture has been done on the basis of the scaling method applied 
to modeling of the power loss in the soft magnetic materials \citep{bib:Sokal}- \citep{bib:Sokal4}. Additionally, two advantages have been achieved. The first one consists of algorithmic structure of the applied procedure which will enable creation of a computer implementation. 
The second one is the case with which SMC's properties can be tailored. As an example the technological parameters for production of a material with less excess losses has been designed and  such a sample has been prepared in accordance with that.

\section{SOFT MAGNETIC COMPOSITE}\label{II}
\begin{figure}[!t]
\centering
\includegraphics[ width=8cm]{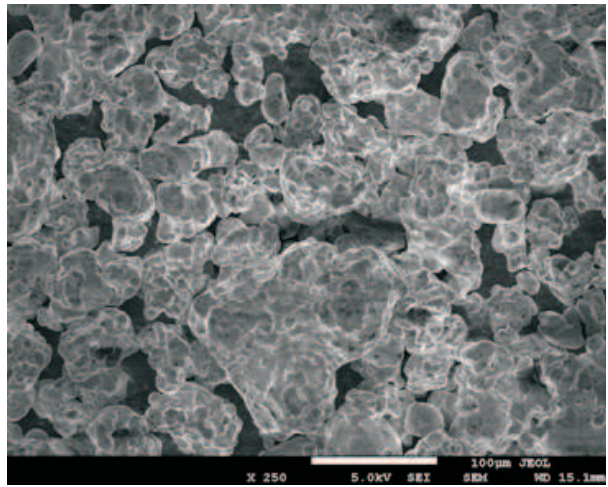}
\caption{Irregular particles of soft magnetic composite} 
\label{Foto.1}
\end{figure}

Soft magnetic composite (SMC) consist generally of ferromagnetic particles distributed in a matrix of binding agent. Metal powder is produced by atomization techniques; a stream of molten metal is atomized by inert gas or water under high pressure, leading to formation of fine liquid metal droplets which then solidify resulting in fine powder particles. The nature of impact media determines shape of powder particles. A jet impacting on water produces irregular particles (Fig.\ref{Foto.1}). In turn, use of the gas as the media produces spherical particles. Grains of iron powder for soft magnetic composites are covered chemically by thin inorganic insulation layer-phosphate glass with thickness of a few nanometers. 
Magnetic composites are prepared by compression moulding; a die is filled with magnetic powder and it is compacted. Then the 'green compact' is cured in a furnace at melting temperature of binding agent. In the case of compression moulding techniques an air atmosphere is used during the curing, frequently.
A commercially available iron powder, produced by water atomization, with a binding agent has been used in the experiment as the soft magnetic composite. It is called Somaloy 500 and is produced by Höganäs AB Company.

\section{Power Loss' Measurements}\label{III}
The samples investigated in this study were produced from commercially available pure iron powders produced by water atomization techniques, with a special surface coating on each and every particle. The powder used in the investigation was Somaloy 500 produced by Höganäs AB Company.
The samples used were ring-shaped with a square cross-section. Samples had the following dimensions, external diameter of 55 mm, internal diameter of 45 mm and thickness of 5 mm.
The power losses Ptot were obtained from measurements of the AC hysteresis cycle according to the IEC Standards 60404-6 using hysteresisgraph AMH-20K-HS produced by Laboratorio Elettrofisico Walker LDJ Scientific. Total power losses Ptot were measured at maximum flux density $B_m$ = 0,1...1,3 T, over a frequency range from 10 to 5000 Hz. During measurements of the total power losses $P_{tot}$ the shape factor of the secondary voltage was equal to $1.111 \pm 1.5 \%$. Measurement error of the total energy losses was equal to $\pm 3 \%$. Measurements of power losses were conducted by recording individual points and the integration area of the hysteresis loop.
Optimization of power loss has been performed with respect to the compacting pressure $p$ and temperature $T$, while the other technological parameters (powder compositions, volume fraction) were constant. This procedure was based on the set of $16$ samples which has been produced according to the measurement matrix of the technological parameters. This matrix is readable from the two first columns of the TABLE \ref{Table:Table1}.
The particle morphology observations were carried out using a scanning electron microscope JSM-7600F produced by Joel.

\section{Power Loss' Modelling and Technological Parameters Values' Desining}\label{IV}
 Soft Magnetic Composites are assumed to be complex system where the function of energy losses obeys the scaling law. This assumption leads to the expression of total energy loss in form of general homogeneous function \cite{bib:Sokal}:
\begin{equation}
\label{general}
P_{tot}(f,B_{m})=B_{m}^{\beta}F\left(\frac{f}{B_{m}^{\alpha}}\right),
\end{equation}
where $F(\cdot)$ is an arbitrary function, $\alpha$ and $\beta$ are scaling exponents. This function depends on features of phenomena to be describe. In order to much $F(\cdot)$ with the eddgy current and  hysteresis losses  we choose for $F(\cdot)$ the power series as a rough description of the energy losses  $P_{tot}$ \citep{bib:Sokal}: 
 \begin{widetext}
 \begin{equation} 
\label {eq8} 
P_{tot}= B_m^{\beta}\,(\Gamma_{1}\,\frac{f}{B_m^{\alpha}}+ \Gamma_{2}\,\left(\frac{f}{B_m^{\alpha}}\right)^2 +%\nonumber\\
\Gamma_{3}\,\left(\frac{f}{B_m^{\alpha}}\right)^3 +\Gamma_{4}\,\left(\frac{f}{B_m^{\alpha}}\right)^4  ), 
\end{equation}
\end{widetext}
%\begin{eqnarray}
%\label {eq9} 
%P_{tot}= B_m^{\beta}\,[\Gamma_{1}\frac{f}{B_m^{\alpha}}+ \Gamma_{2}(\frac{f}{B_m^{\alpha}})^2 +... ], 
%\end{eqnarray}
where $f$ - frequency, $B_{m}$ - amplitude of magnetic field's induction. 
Values of $\alpha$, $\beta$ and amplitudes $\Gamma_{n}$  have been estimated for different values of pressure and temperature. It is easy to recognize by the powers of $f$ that the two first terms in (\ref{eq8}) correspond to hysteresis losses $P_{h}$ and eddy current losses $P_{c}$, whereas higher terms correspond to excess losses $P_{ex}$, respectively:
\begin{equation}
\label{Ptot}
P_{tot}= P_{h}+P_{c}+P_{ex}.
\end{equation}
For more details concerning this separation formula and scaling see to \citep{bib:Sokal4}.

\subsection{Energy losses' data of SMC and the scaling law}

The parameters' values of (\ref{eq8}) has been estimated for each sample's measurement data by minimization of $\chi^2$
\begin{equation}
\label{Chi2}
\chi^{2}=\Sigma_{i=1}^{n}\left(\frac{P_{tot\,i}-P_{exp \,i}}{B_{m\,i}^{\beta}}\right)^{2}
\end{equation}
 using the simplex method of Nelder and Mead \citep{bib:Nelder}. 

\begin{center}
%\begin{group}
%\squeezetable
\begin{table*}%[!t]
\renewcommand{\arraystretch}{1.3}
\caption{ Somaloy 500. Values of scaling exponents and coefficients of (\ref{eq8}) v.s. hardening temperature and  pressing pressure 
\label{Table:Table1}}
\begin{tabular}{|c|c||c|c|c|c|c|c|}
\hline
Temp.[${}^{o}{C}$] & Pressure [MPa] & $\alpha$ & $\beta$ & $\Gamma_{1}[\frac{m^{2}}{s^{2}}T^{\alpha-\beta}]$ & $\Gamma_{2}[\frac{m^{2}}{s}T^{2\alpha-\beta}]$ & $\Gamma_{3}[m^{2}T^{3\alpha-\beta}]$ & $\Gamma_{4}[{m^{2}}{s}\,T^{4\alpha-\beta}]$ \\
\hline\hline
500&500 & -1,312223 &	-0,01052897	& 0,170825 &	3,60591E-05	& 1,95309E-08	& -2,25448E-12\\
500&600 & -1,383117 &	-0,1249742 & 	0,1525481 &	3,32813E-05 &	9,25383E-09 &	-1,17698E-12\\
500&700 & -1,73502 &	-0,5173269 & 	0,1558437 &	2,39335E-05	& 2,30877E-09 &	-8,07475E-14\\
500&900& -1,394503 &	-0,082346  &	0,1012103 &	6,06463E-05	& -8,03122E-09 &	7,87711E-13\\
400 &800&-1,47298 &-0,028389 &	0,1831606	& 1,34725E-05 & 3,68924E-09 & -1,18518E-13 \\
450 &800&-1,596151&-0,1231805&0,1493641&2,481657E-05&-1,217578E-09&6,120261E-14\\
550&800&-2,034396&-1,325493&0,1055319&1,40738E-04&-1,06631E-08&4,54085E-13\\
600  &800& -1,608202&	-0,2323374&	1,219838	&0,000894116&	-5,30152E-08&	-1,66434E-11	\\
438 & 764 & -2,059196  & -1,014382 & 0,1395258 & 7,941049E-05 & -1,446713E-08 & 1,482247E-12\\
469 & 764 & -1,402494 & -0,0137938 & 0,1340093 & 5,373893E-05 & -6,098702E-09 & 6,883649E-13\\
531 & 764 & -3,09792 & -2,318892 & 0,1293812 & 6,054176E-05 & -3,967688E-09 & 2,019266E-13\\ 
480 & 780 &-1,634466&-0,3416979&0,1242102&6,852739E-05&-9,746355E-09&1,001089E-12\\
\hline 					
\end{tabular}\\ \vspace{1mm}
\end{table*}
\end{center}
% \end{widetext}

\begin{figure}%[!t]
\centering
\includegraphics[ width=8cm]{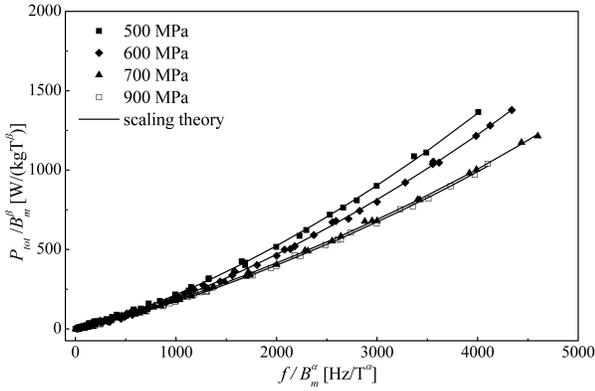}
\caption{Somaloy 500. $P_{tot}/B_{m}^{\beta}$ vs. $f/B_{m}^{\alpha}$ for $T=450^{o}C$, $p=500,600,700$ and $900 MPa.$. The continuous lines correspond to (\ref{eq8}).} 
\label{Fig.3}
\end{figure}

\begin{figure}%[!t]
\centering
\includegraphics[ width=8cm]{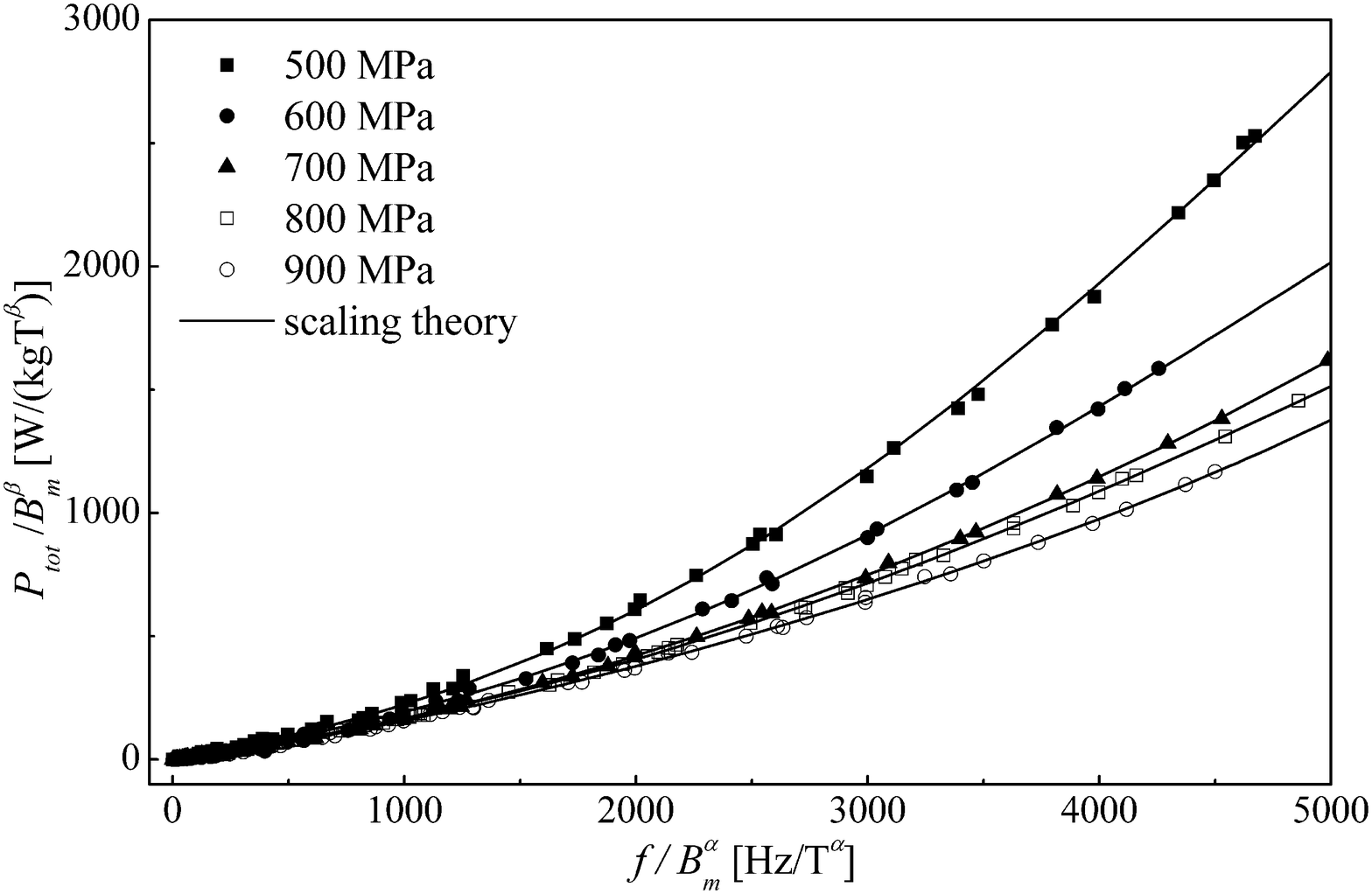}
\caption{Somaloy 500. $P_{tot}/B_{m}^{\beta}$ vs. $f/B_{m}^{\alpha}$ for $T=500^{o}C$, $p=500,600,700,800$ and $900 MPa$. The continuous lines correspond to (\ref{eq8}).} 
\label{Fig.2}
\end{figure}

\begin{figure}%[!t]
\centering
\includegraphics[ width=8cm]{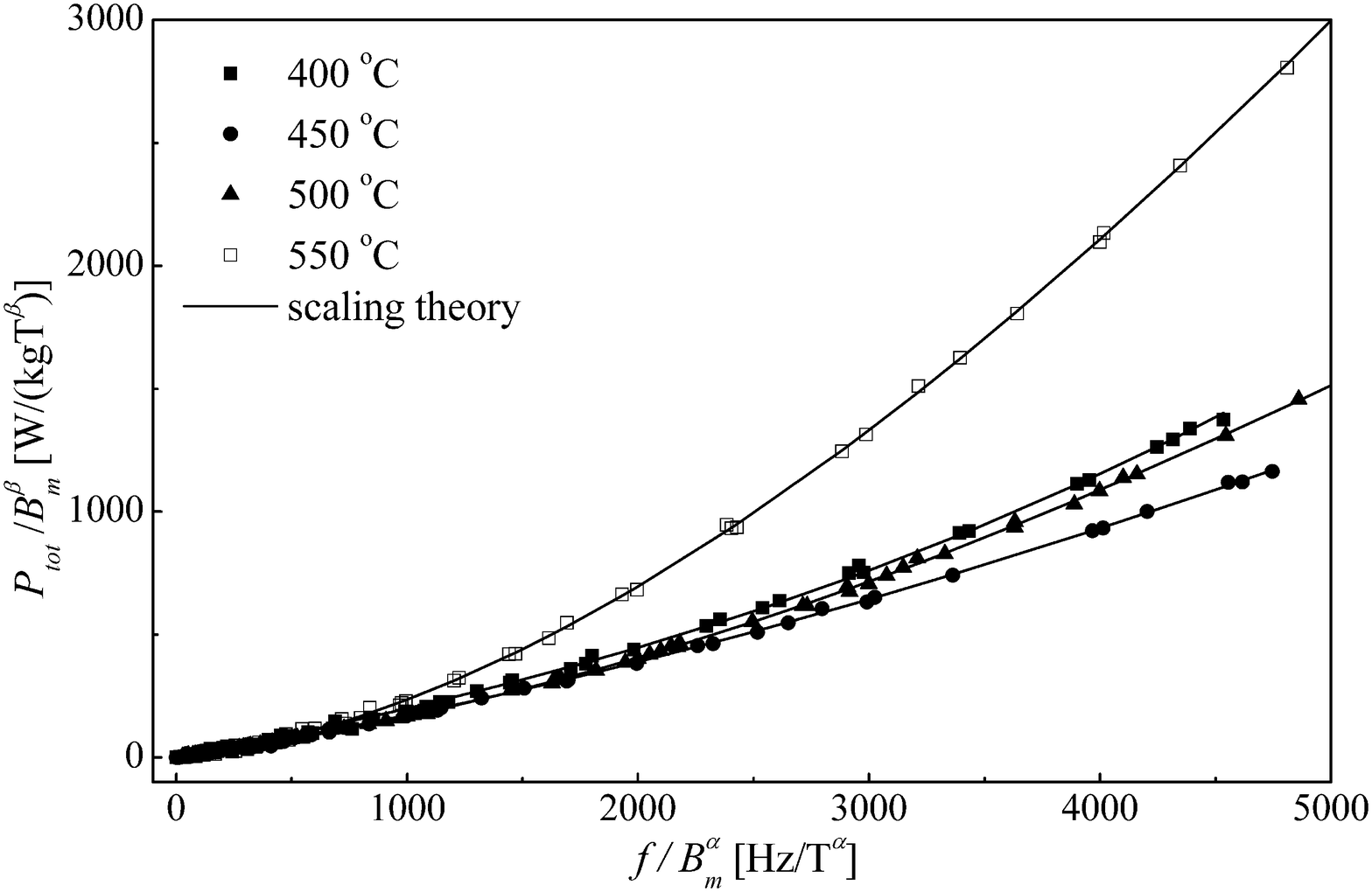}
\caption{Somaloy 500. $P_{tot}/B_{m}^{\beta}$ vs. $f/B_{m}^{\alpha}$ for $p=800 MPa$, $T=400,450,500$ and $550^{o}C$.  The continuous lines correspond to (\ref{eq8}).} 
\label{Fig.4}
\end{figure}

\begin{figure}%[!t]
\centering
\includegraphics[ width=8cm]{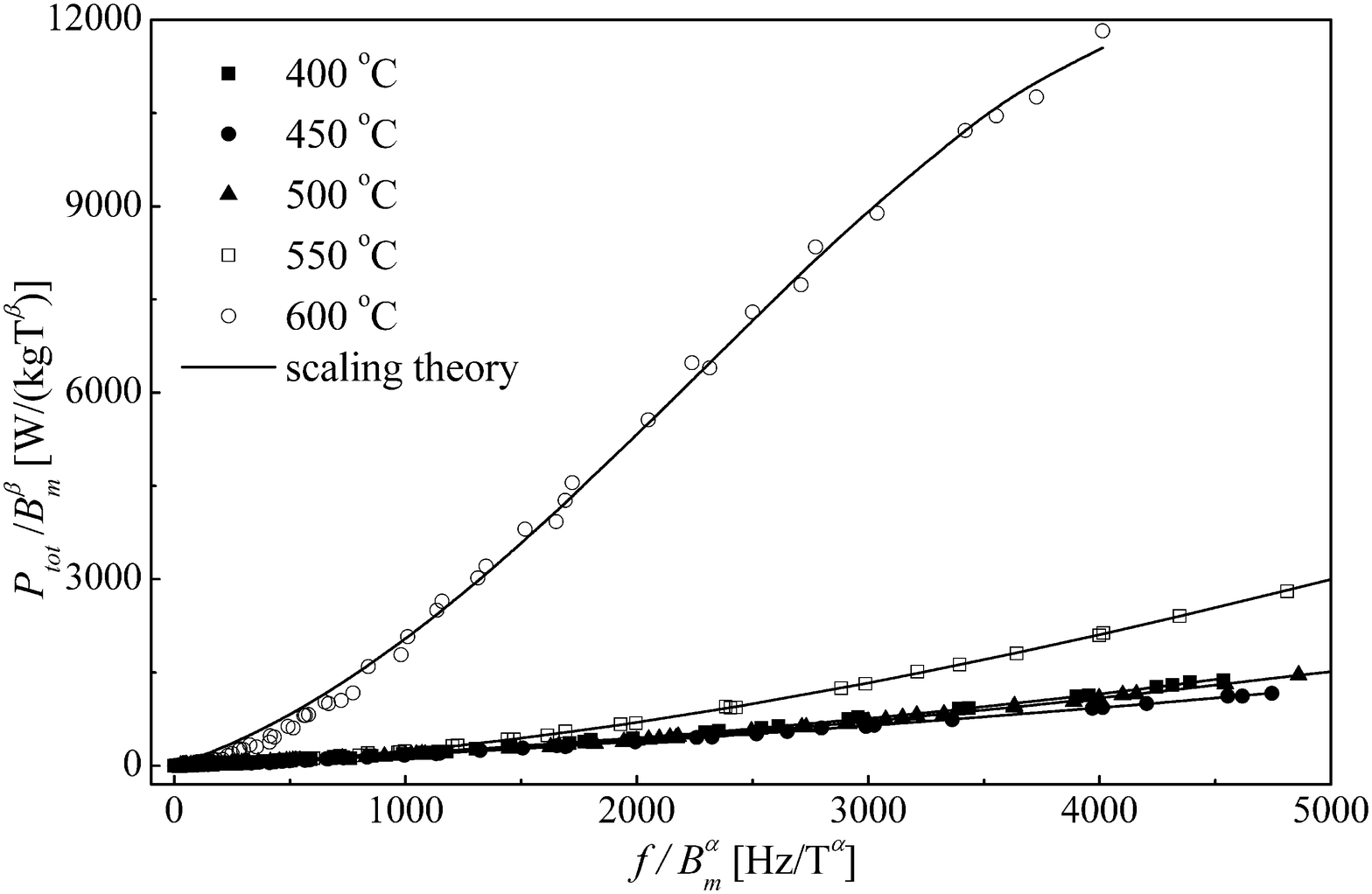}
\caption{Somaloy 500. $P_{tot}/B_{m}^{\beta}$ vs. $f/B_{m}^{\alpha}$ for $p=800MPa$, $T=400,450,500,550$ and $600^{o}C.$ The continuous lines correspond to (\ref{eq8}).} 
\label{Fig.5}
\end{figure}

For results see TABLE \ref{Table:Table1}. Using these values and (\ref{eq8}) and the estimators of energy losses in considered samples has been obtained. Each continuous line represents the corresponding model predictions. Comparisons of estimations and the measurement results  were depicted in Fig.\ref{Fig.2} and Fig. \ref{Fig.3} for $T=constant$ and in Fig.\ref{Fig.4} and Fig. \ref{Fig.5} for p= constant. Basing on these results one can derive the following  conclusions:\\
\begin{itemize}
\item
The energy losses' experimental data obey the scaling law (\ref{general}) which means that $P_{tot}/B_{m}^{\beta}$ depends only on the one effective variable $\tilde{f}=\frac{f}{B_{m}^{\alpha}}$. This property enables to present all measurement results for $P_{tot}$ v.s. $f$ and $B_{m}$ on a one-dimensional curve instead of the two-dimensional surface, see Fig.\ref{Fig.3}-Fig.\ref{Fig.5}.
\item
The power loss characteristics $T=constant$ for different compacting pressures  do not cross each other except the origin point                                $(\tilde{P}_{tot}=0,\frac{f}{B_{m}^{\alpha}}=0)$ , and similarly  the power loss characteristics $p=constant$ for different compacting temperatures  do not cross each other except the mentioned origin, see Fig.\ref{Fig.3}-Fig.\ref{Fig.5}.
\item
Let $\tilde{P}_{tot,T_{j},p_{k}}(\tilde{f})=P_{tot,T_{j},p_{k}}(\tilde{f})/B_{m}^{\beta}$ be scaled total energy loss in sample material which has been compacted in $T_{j}$ temperature and under $p_{k}$ pressure. Let $T_{1}=T_{2}$,  then $\exists p_{c}(T)$ such that $\forall \tilde{f}\ge 0$  and $\forall p_{c}(T_{1})\ge p_{1}>p_{2}$.the following relation holds: $\tilde{P}_{tot,T_{1},p_{1}}(\tilde{f})<\tilde{P}_{tot,T_{1},p_{2}}(\tilde{f})$,  see Fig.\ref{Fig.3}-Fig.\ref{Fig.2}.
\item
Let $\tilde{P}_{tot,T_{j},p_{k}}(\tilde{f})=P_{tot,T_{j},p_{k}}(\tilde{f})/B_{m}^{\beta}$ be scaled total energy loss in sample material which has been compacted in $T_{j}$ temperature and under $p_{k}$ pressure.  Let $p_{1}=p_{2}$. Then $\exists T_{c}(p)$ such that $\forall \tilde{f}\ge 0$ and $\forall T_{c}(p) \ge T_{1}>T_{2}$ the following relation holds: $\tilde{P}_{tot,T_{1},p_{1}}(\tilde{f})<\tilde{P}_{tot,T_{2},p_{1}}(\tilde{f})$,  see Fig.\ref{Fig.4}-Fig.\ref{Fig.5}.
\end{itemize}

\begin{figure}%[!t]
\centering
\includegraphics[ width=8cm]{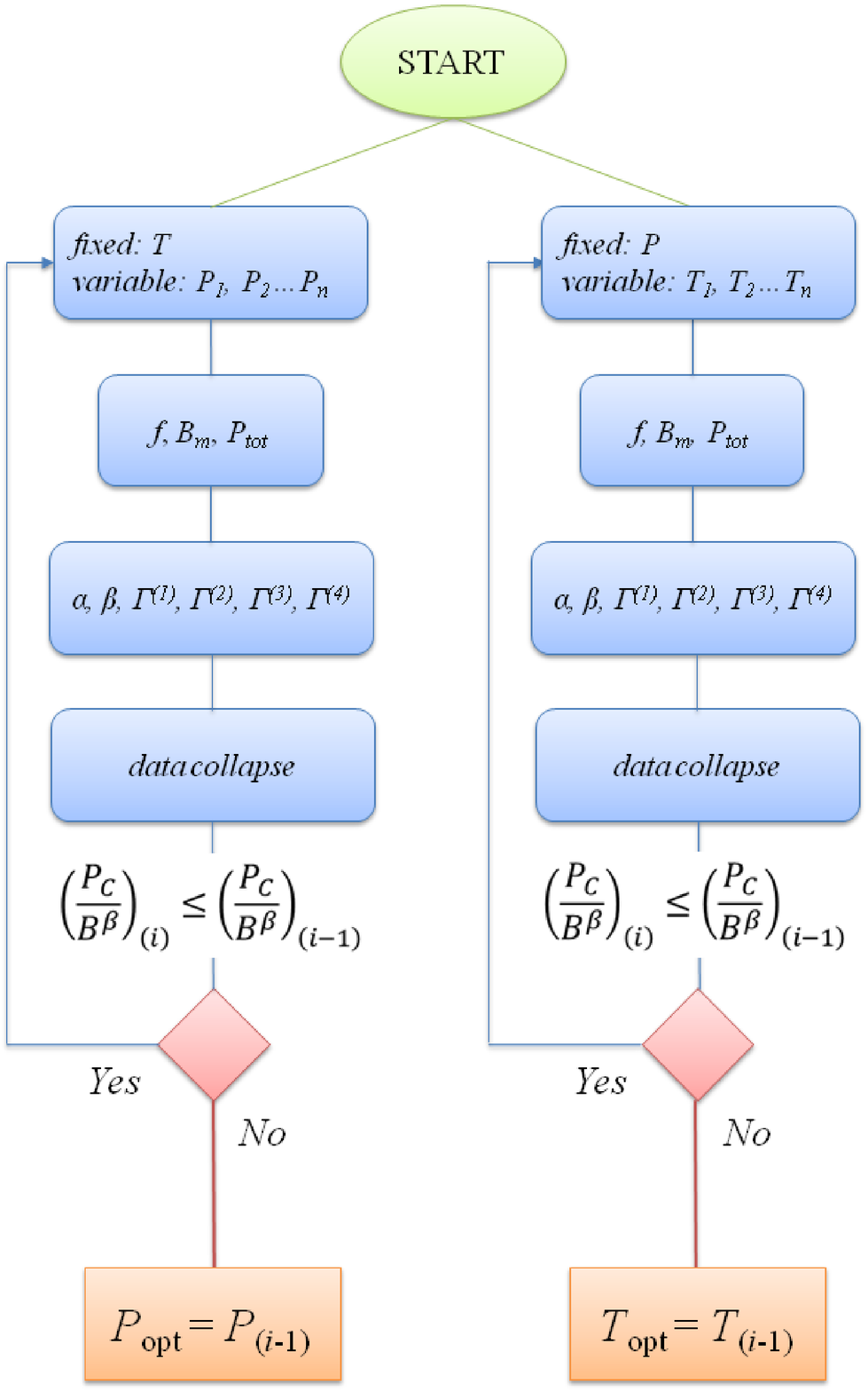}
\caption{Algorithm for designing the best values of technological parameters.} 
\label{Fig.8}
\end{figure}

The data presented here of measured properties and the mathematical model (\ref{eq8}) constitute algorithmic structure. Moreover,  they enable one to optimize the power losses globally for every values of independent variables $f$ and $B_{m}$, therefore for the optimization process their values need not to be assigned.
Graph presenting working algorithm is presented in Fig.\ref{Fig.8} \citep{bib:Jank}.

%\subsection{Asimptotic properties of the energy losses in SMC v.s. \tilde{f}}
\section{Desining excess power loss-less material}

\begin{figure}
\centering
\includegraphics[ width=8cm]{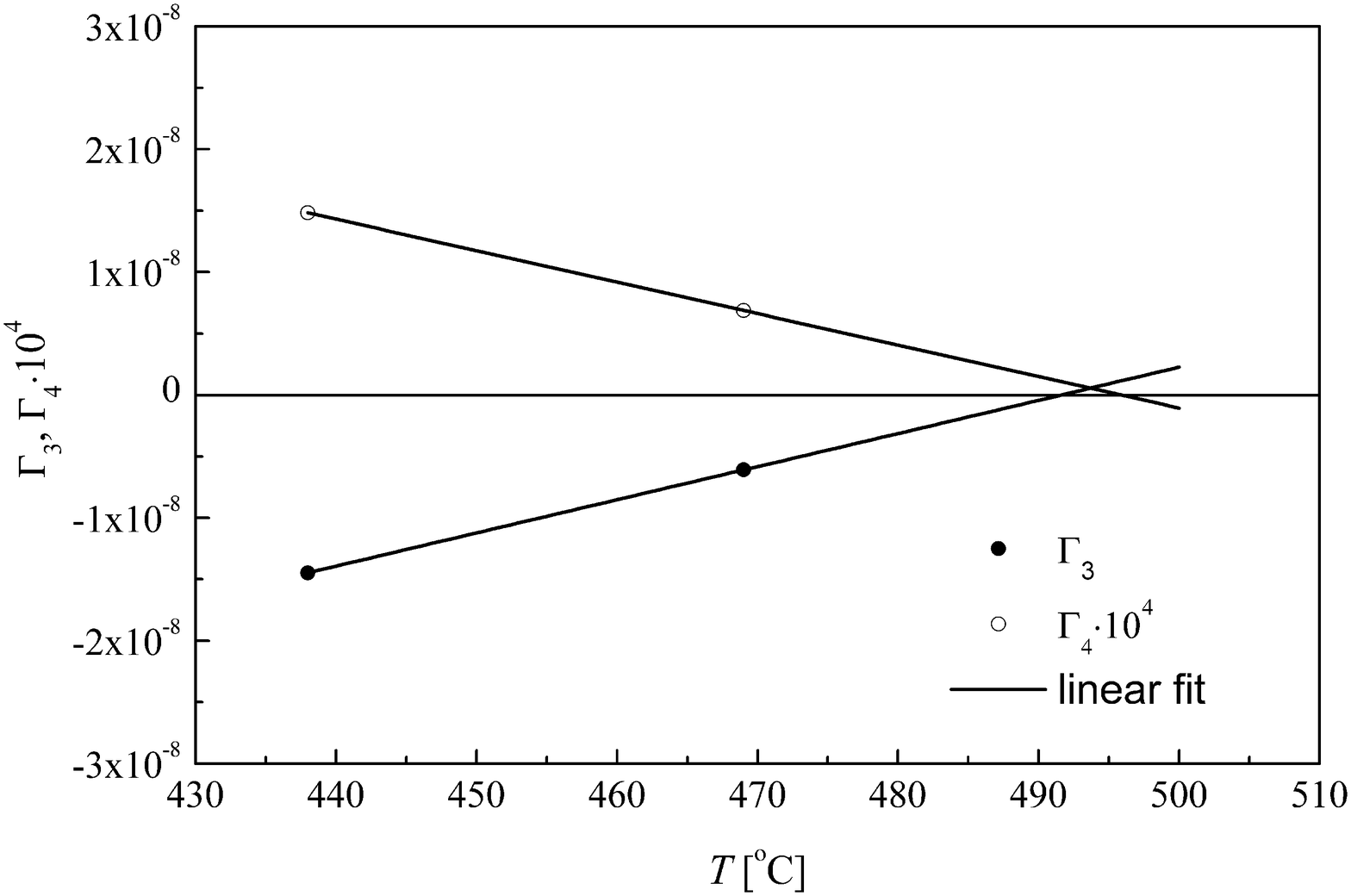}
\caption{Somaloy 500, Interpolation of the technical critical point $T_{tc}=438^oC, p_{tc}=800MPa$.} 
\label{Fig.11}
\end{figure}

\begin{figure}
\centering
\includegraphics[ width=8cm]{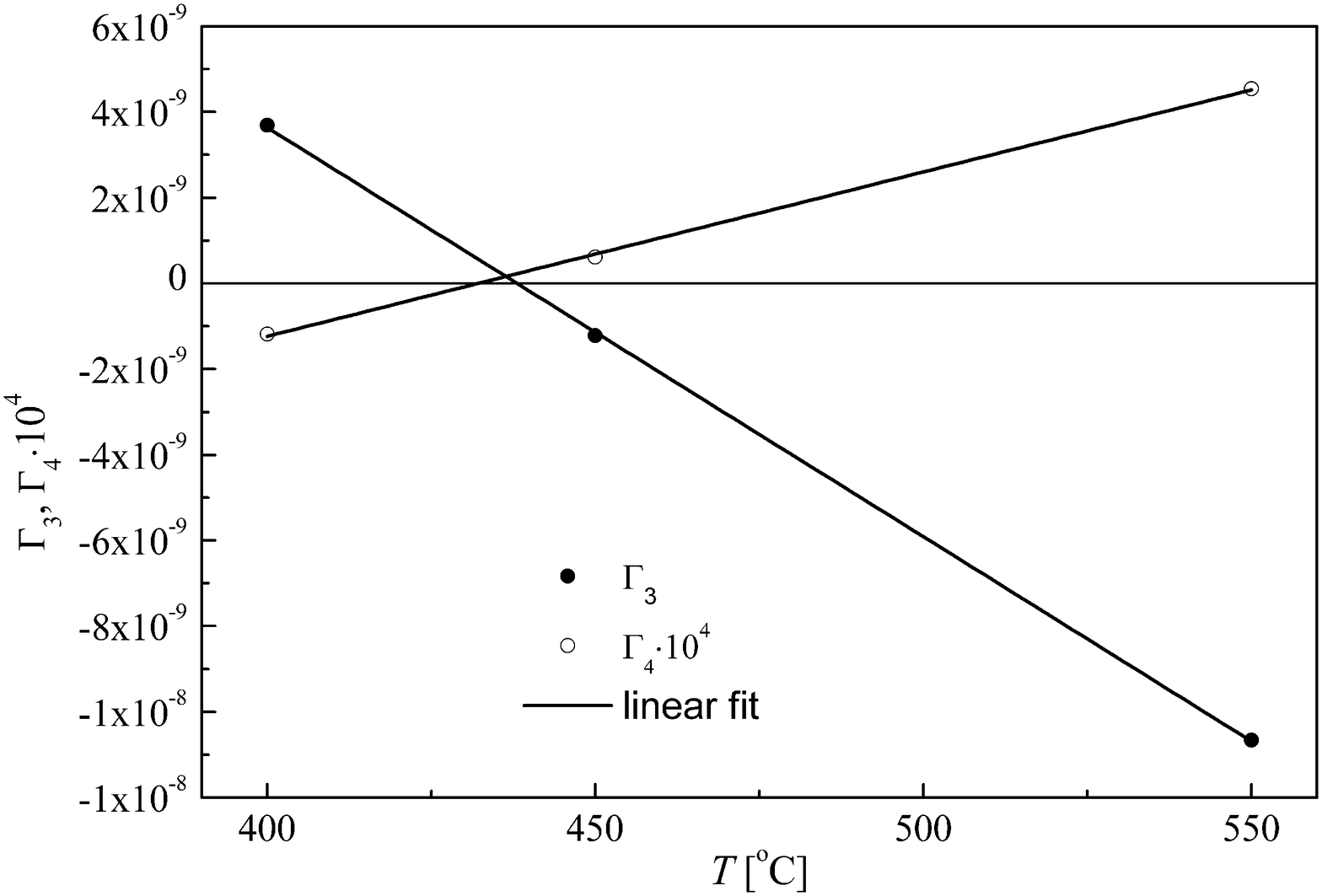}
\caption{Somaloy 500. Extrapolation of the technical critical point $T_{tc}=495^oC, p_{tc}=764MPa$}
\label{Fig.12}
\end{figure}

\begin{figure}
\centering
\includegraphics[ width=8cm]{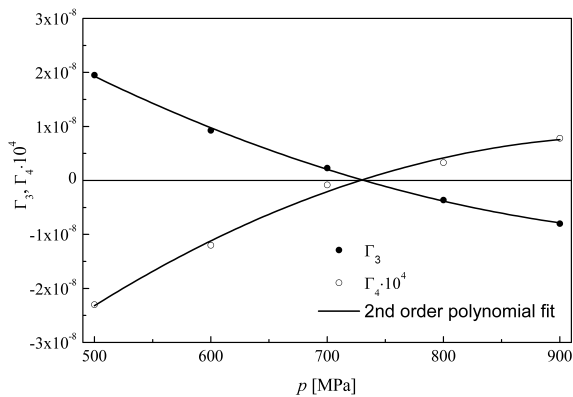}
\caption{Somaloy 500. Interpolation of the technical critical point $T_{tc}=500^oC, p_{tc}=728MPa$} 
\label{Fig.13}
\end{figure}

\begin{figure}
\centering
\includegraphics[ width=8cm]{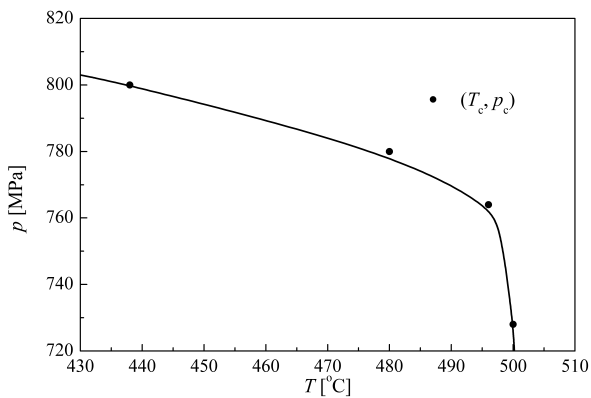}
\caption{Somaloy 500. Technical critical curve in the $T,p$ plane.\\
 Markers correspond to the points determined  in this paper.} 
\label{Fig.14}
\end{figure}

Based on the results of \citep{bib:Sokal4} and on the subsection \ref{IV} of this paper it is possible to design and produce a SMC's material which does not exhibit the excess contribution to the total power loss, e.g.: $P_{tot}=P_{h}+P_{c}$. Let us call the values of technical parameters leading to such a material technical critical parameters: $T_{tc}$ and $p_{tc}$. Let us truncate formula (\ref{eq8}) above the second order term. The obtained form described materials with less excess power loss: 
\begin{equation}
\label{trunk1}
P_{tot}= B_m^{\beta}\,\left(\Gamma_{1}\,\frac{f}{B_m^{\alpha}}+ \Gamma_{2}\,\left(\frac{f}{B_m^{\alpha}}\right)^2\right).
\end{equation} 
Comparing (\ref{eq8}) and (\ref{trunk1}) one can conclude that materials with less excess power loss are characterized by $\Gamma_{3}=0$ and $\Gamma_{4}=0$. Therefore, in order to satisfy the formulated task one should produce a sample under the technological conditions leading to these values of $\Gamma_{3}$ and $\Gamma_{4}$, e.g. $T_{tc}$ and $p_{tc}$. How to do it? Having at least two samples which have been produced in different points $(T,p_{1}),\, (T,p_{2})$ or $(T_{1},p), \,(T_{2},p)$  one can approximate  the required values of $T_{tc}$ and $p_{tc}$ by inter or extrapolation. Fig.\ref{Fig.11} - Fig.\ref{Fig.13} presents three examples of such calculations. The best accuracy has been achieved for the example presented in Fig.\ref{Fig.13}. Note that both curves $\Gamma_{3}=\Gamma_{3}(T=constant,p)$ and $\Gamma_{4}=\Gamma_{4}(T=constant,p)$ get $0$ for the same values of $p$. By this manner the technological critical parameters have been determined for three materials with less excess power loss (see TABLE \ref{Table:Table2}). These  points are sufficient to draw a curve in the $(T,p)$ plane (see Fig.\ref{Fig.14}) which is approximation representation of the whole set of technological parameters values leading to materials with less excess power loss. Note that this curve divides $T,p$ space into two parts. The upper part consists of better values of $T$ and $p$ which lead to smaller values of the power loss in comparison to the lower part. However, this space is bounded from above, both from $p$ and $T$ sides (see Fig.\ref{Fig.2}-Fig.\ref{Fig.5}). This revelation helps to limit $T,p$ space to a small subset whose boundaries are determined by the set of the best values of technological parameters.
%\begin{figure}
%\includegraphics[ width=8cm]{500MPax3.eps}
%\label{Fig.9}
%\end{figure} 
%\begin{figure}
%\centering
%\includegraphics[ width=8cm]{900MPax3.eps}
%\caption{Somaloy 500.b.} 
%\label{Fig.10}
%\end{figure} 
  

\begin{center}
%\begin{group}
%\squeezetable
\begin{table*}%[!t]
\renewcommand{\arraystretch}{1.3}
\caption{  Values of  hardening temperature and  pressing pressure corresponding to excess power loss less materials
\label{Table:Table2}}
\centering
\begin{tabular}{|c||c|c|c|c|}
\hline
$T_{tc}[{}^{o}{C}]$ &438 &480 &495 &500\\ \hline
$p_{tc} [MPa]$    &800 &780 &764 &728\\
\hline 			
\end{tabular}\\ \vspace{1mm}
\end{table*}
\end{center}

\section{Conclusions}
Modern engineering of SMM  designs materials which posses required physical and technological properties. Particularly, technology of SMC enables one to optimize both, the power loss and the magnetic properties.  The results presented here concern only the power losses and they are part of the Complex researching programme (No. N N507 249940) which includes investigations and optimization also that ones. However, in the case of this paper's topic the following results have been achieved. Power losses have been measured in Somaloy 500 samples for a wide range of frequency and magnetic induction. These samples have been prepared under a wide range of the hardening temperatures and pressures. The power loss characteristics have been derived by assuming that investigated samples obeyed the scaling law. Agreement obtained between experimental data and the scaling theory has confirmed that the assumptions of this paper are true. Moreover, the experimental data of the given sample have been collapsed to a single curve which represented  measurements for all values of $f$ and $B_{m}$. Therefore, the the scaling transforms the losses characteristics from the two dimensional surfaces to the one dimensional curves. The samples were produced according two methods: for different pressures  with $T=constant$  and at different temperatures with $p=constant$. In both cases the power losses decrease with increasing pressure and with increasing temperature. These trends in decreasing losses stopped for certain critical values $p_{c}(T)$ and $T_{c}(p)$, respectively. Above these values the power losses increase suddenly. Therefore,  $p_{c}(T)$ and $T_{c}(p)$ are sought after solutions for optimal values of pressure and temperature. The formula of presented power losses consists of four terms. According to \citep{bib:Sokal4} the two first terms in (\ref{eq8}) correspond to $P_{h}$ and $P_{c}$. However, the sum of the third and fourth terms describes $P_{ex}$.  By the appropriate choice of $p$ and $T$ one can reduce $P_{ex}$ to zero. In this way it is possible to design materials with less excess losses. Three such points have been determined experimentally and a curve was drawn representing all Somaloy 500 materials with less excess losses in the $(p,T)$ plane. The obtained curve divides $(p,T)$ plane into subsets of sample high and low losses.

 Finally we hope to achieve compact algorithm enabling the design of both magnetic and power loss properties.

%\begin{center}
%'\begin{group}
%'\squeezetable
%\begin{table}[!t]
%\renewcommand{\arraystretch}{1.3}
%\caption{Ranges of measured magnitudes}
%\label{Table1}
%\centering
%\begin{tabular}{|c||c|c|c|}
%\hline
% Sample & $f[Hz]$ & $B_{m}[T]$ & $P_{tot}[\frac{W}{kg}]$ \\
%\hline\hline
%$P{1}$ & 10-400 & 0.101-1.201 &0.001-2.666 \\
%$P{2}$ & 10-400 & 0.100-0.999 &0.001-1.755	\\
%$P{4}$ & 1-500 & 0.116-1.800&0.000-74.519\\
%$P{7}$ & 40-400 & 0.100-0.700 &	0.002-2.427\\
%\hline 					
%\end{tabular}\\ \vspace{1mm}
%\end{table}
%'\end{group}
%\end{center}
%'\end{widetext}

%\begin{figure}[!t]
%\centering
%\includegraphics[ width=8cm]{Graph2_1b.eps}
%\caption{$P_{tot\,1,2}$ vs. $f_{1,2}$ for P2 - amorphous alloy 
% $\textrm{Co}_{71.5}\textrm{Fe}_{2.5}\textrm{Mn}_{2}\textrm{Mo}_{1}\textrm{Si}_{9}\textrm{B}_{14}$} 
%\label{Fig.2}
%\end{figure} 

\bibliographystyle{plainnat}
% \begin{widetext}

\end{document}